\documentclass[preprint,preprintnumbers, prd, floatfix, superscriptaddress,nofootinbib] {revtex4-1}
\usepackage{epsfig}
\usepackage{subfigure}
\usepackage{dcolumn}
\usepackage{bm}
\usepackage[usenames ,dvipsnames]{xcolor}
\usepackage{slashed}
\usepackage{graphicx,color}
\begin{document}
\title{Distinguishing (Dirac or Majorana) neutrinos in  purely leptonic decays of leptons}

\author{Yao Yu}
\email{Corresponding author: yuyao@cqupt.edu.cn}
\affiliation{Chongqing University of Posts \& Telecommunications, Chongqing, 400065, China}
 \affiliation{Department of Physics and Chongqing Key Laboratory for Strongly Coupled Physics, Chongqing University, Chongqing 401331, People's Republic of China}
\author{Bai-Cian Ke}
\email{Corresponding author:baiciank@ihep.ac.cn}
\affiliation{Zhengzhou University, Zhengzhou 450001, People's Republic of China}

\date{\today}

\begin{abstract}
  We investigate purely leptonic decays of leptons
  $\l^{\prime-}\to l^-\bar{\nu}_l\nu_{l^{\prime}}$ to distinguish Dirac or
  Majorana neutrinos. We derive the differences of the decay width (and
  associated quantities) between the two neutrino hypotheses by the effect of
  identical particles. Evidence from experimental data makes the hypothesis of
  Majorana neutrinos very unlikely and our results strongly favor the
  hypothesis of Dirac neutrinos (in the standard three-neutrino theory).
  Moreover, our exploration for decay widths can be extended to the
  polarization angles of leptons, which will inspire new information for
  experimental and theoretical studies.
\end{abstract}

\maketitle
\section{introduction}
The question of whether neutrinos are Dirac~\cite{Dirac:1928hu} or
Majorana~\cite{Majorana:1937vz} particles lies at the heart of particle
physics. Experimentally, neutrinoless double beta decay~\cite{Schechter:1981bd}
is the most popular access to test the hypothesis that neutrinos are Majorana
particles. Unfortunately, it only keeps setting higher precision limits so
far~\cite{Adams:2022hji}, and whether neutrinos being Majorana particles or not
are not concluded yet. This leads to an urgent need to find another way to
identify neutrinos (Dirac or Majorana). Fortunately, we have found that
purely leptonic decays of leptons provides an ideal opportunity to identify
neutrinos considering the precise experimental results of lepton decays.

In this work, we investigate purely leptonic decays of leptons
$\l^{\prime-}\to l^-\bar{\nu}_l\nu_{l^{\prime}}$, where
$(l^{\prime}=\tau\,,l=e,\mu)$ or $(l^{\prime}=\mu\,,l=e)$.
According to neutrino oscillations~\cite{Pontecorvo:1957qd,Maki:1962mu},
$\l^{\prime-}\to l^-\bar{\nu}_l\nu_{l^{\prime}}$ is the superposition of
several $\l^{\prime-}\to l^-\bar{\nu}_i\nu_{j}$, where $\bar{\nu}_i$ and
$\nu_{j}$ indicate the mass eigenstates of neutrinos. It makes no different
between Dirac and Majorana neutrino hypotheses to investigate purely leptonic
decays of leptons in their flavor eigenstates,
i.e.~$\l^{\prime-}\to l^-\bar{\nu}_i\nu_{j}$ $(i\neq j)$, but it has obvious
distinction between the two hypotheses to investigate in neutrinos' mass
eigenstates, i.e.~$\l^{\prime-}\to l^-\bar{\nu}_i\nu_{i}$ when $(i=j)$.
$\bar{\nu}_i,\nu_{i}$ are not identical particles in Dirac neutrinos, while
identical in Majorana neutrinos. Analogous to what mentioned
in~\cite{Cheng:2007si,Cheng:2022vbw}, the branching ratio of
$D^+\to \pi^+\pi^-\pi^+$ must include the influence of two identical $\pi^+$'s.
Identical neutrinos bring an extra interference term than non-identical
neutrinos do, which causes the decay width of purely leptonic decays of leptons
under Majorana neutrino hypothesis always narrower than that under Dirac
neutrino hypothesis. The key reason is Fermi-Dirac statistics for identical
neutrinos (details will be shown in Formalism).
 
Besides having a pair of neutrino and anti-neutrino, which could be identical,
in the final states, the purely leptonic decays of leptons
$\l^{\prime-}\to l^-\bar{\nu}_l\nu_{l^{\prime}}$ can play as gold decay
channels to distinguish Dirac or Majorana neutrinos due to two other reasons.
Firstly, the neutrino oscillations have a substantial flavour mixing angle,
which means different flavour neutrinos seriously overlap in the neutrino mass
eigenstate space, and then cause a large proportion of identical particles
under Majorana neutrino hypothesis. This will make a significant difference of
conclusions between the two neutrino hypotheses. Secondly, the purely leptonic
decays of leptons have no strong or electromagnetic interaction, which makes
them can be calculated very accurately and precisely even in four-fermion
interactions. The small theoretical uncertainty leaves no ambiguity to draw
out the difference of the two neutrinos hypotheses.

At last, although we investigate only the decay width (and associated
quantities), polarization angles of $l^{\prime-}$ and $l^-$ are also excellent
places to distinguish Dirac or Majorana neutrinos, which will inspire new
information for experimental or theoretical studies. Investigating
polarization angles in purely leptonic decays of leptons is the next step of
our work plan .

\section{Formalism}
The decay widths of the $\l^{\prime-}\to l^-\bar{\nu}_l\nu_{l^{\prime}}$ decay,
considering neutrino oscillation in the standard three-neutrino theory, is
given by
\begin{eqnarray}
\Gamma(\l^{\prime-}\to l^-\bar{\nu}_l\nu_{l^{\prime}})=\sum^{i=3}_{i=1}\sum^{j=3}_{j=1}\Gamma(\l^{\prime-}\to l^-\bar{\nu}_i\nu_{j})\,.
\end{eqnarray}

At first, we assume neutrinos are Dirac particles, 
the amplitude of $\l^{\prime-}\to l^-\bar{\nu}_i\nu_{j}$ can be written as
\begin{eqnarray}\label{amp1a}
{\cal M_D}(\l^{\prime-}\to l^-\bar{\nu}_i\nu_{j})&=&\frac{G_F}{\sqrt 2}U_{li}U^*_{l^{\prime}j}
\bar{u}_l\gamma^\mu(1-\gamma^5)\nu_i\bar{u_j}\gamma_\mu(1-\gamma^5)u_{l^\prime}\,,
\end{eqnarray}
where $U_{li}$ and $U^*_{l^{\prime}j}$ are the matrix elements of PMNS
matrix~\cite{Maki:1962mu}. In the ${l^\prime}$ rest frame, the partial
decay width reads
\begin{eqnarray}\label{er}
  \Gamma_{\cal D}(\l^{\prime-}\to l^-\bar{\nu}_i\nu_{j}) &=&\frac{1}{(2\pi)^3}\frac{1}{32m^3_{l^\prime}}\int\overline{|{\cal M_D}|^2} dm^2_{12} dm^2_{23}\nonumber\\
  &=&|U_{li}|^2|U^*_{l^{\prime}j}|^2\frac{G^2_F m^5_{l^{\prime}}}{24(2\pi)^3}K_{ll^{\prime}}
\end{eqnarray}
with $m^2_{12}=(p_l+p_{\nu_j})^2$, $m^2_{23}=(p_l+p_{\bar{\nu}_i})^2$,
$K_{ll^{\prime}}=1-8y_{ll^{\prime}}+8y^3_{ll^{\prime}}-y^4_{ll^{\prime}}-12y^2_{ll^{\prime}}\ln y_{ll^{\prime}}$,
and $y_{ll^{\prime}}= m^2_{l}/ m^2_{l^{\prime}}$. With knowing of PMNS
matrix is unitary, we can write the partial decay width as~\cite{Tsai:1971vv}
\begin{eqnarray}
\Gamma_{\cal D}(l^{\prime-}\to l^-\bar{\nu}_l\nu_{l^{\prime}})&=&\frac{G^2_F m^5_{l^{\prime}}}{24(2\pi)^3}K_{ll^{\prime}}\,.
 \end{eqnarray}
Then we reach the decay width $\Gamma(\tau^{-}\to \mu^-\bar{\nu}_\mu\nu_{\tau})$ versus the decay width $\Gamma(\tau^{-}\to e^-\bar{\nu}_e\nu_{\tau})$:
\begin{eqnarray}
  {\cal R_{D}} &=& \frac{\Gamma_{\cal D}(\tau^{-}\to \mu^-\bar{\nu}_\mu\nu_{\tau})}{\Gamma_{\cal D}(\tau^{-}\to e^-\bar{\nu}_e\nu_{\tau})}
  =\frac{K_{\mu\tau}}{K_{e\tau}}
\end{eqnarray}

On the other side, using the same way in the assumption of Dirac neutrinos, the
partial decay width with the assumption of Majorana neutrinos
$\Gamma_{\cal M}(\l^{\prime-}\to l^-\bar{\nu}_i\nu_{j})$ in the case of
$(i\neq j)$ is given by
\begin{eqnarray}\label{FF1_2}
\Gamma_{\cal M}(\l^{\prime-}\to l^-\bar{\nu}_i\nu_{j})
&=&
\Gamma_{\cal D}(\l^{\prime-}\to l^-\bar{\nu}_i\nu_{j})\,,  \,\,\,\,(i\neq j)
\end{eqnarray}
but it is a quiet different story for
$\Gamma_{\cal M}(\l^{\prime-}\to l^-\bar{\nu}_i\nu_{i})$ in the case that
$\bar{\nu}_i,\nu_{i}$ are identical, where the effects of Fermi-Dirac statistics
must be taken into account. As similar way in Ref.~\cite{Cheng:2007si,Cheng:2022vbw}
(in which the effects of Bose-Einstein statistics is taken into account),
the amplitude of $\l^{\prime-}\to l^-\bar{\nu}_i\nu_{i}$ is given by
\begin{eqnarray}
  {\cal M_M}(\l^{\prime-}\to l^-\bar{\nu}_i\nu_{i})&=&\frac{G_F}{\sqrt 2}U_{li}U^*_{l^{\prime}j}\{\bar{u}_l(p_3)\gamma^\mu(1-\gamma^5)\nu_i(p_1)\bar{u_i}(p_2)\gamma_\mu(1-\gamma^5)u_{l^\prime}(p_0)\nonumber\\
  &-&\bar{u}_l(p_3)\gamma^\mu(1-\gamma^5)\nu_i(p_2)\bar{u_i}(p_1)\gamma_\mu(1-\gamma^5)u_{l^\prime}(p_0)\}.
\end{eqnarray}
The partial decay width then reads
\begin{eqnarray}\label{FF1}
\Gamma_{\cal M}(\l^{\prime-}\to l^-\bar{\nu}_i\nu_{i})
&=&
\frac{1}{2}\frac{1}{(2\pi)^3}\frac{1}{32m^3_{l^\prime}}\int\overline{|{\cal M_M}(\l^{\prime-}\to l^-\bar{\nu}_i\nu_{i})|^2} dm^2_{12} dm^2_{23}\,,
\end{eqnarray}
where the factor of $1/2$ accounts for the effect of Fermi-Dirac statistics.
With the helping of Eqs.~(\ref{er})~and~(\ref{FF1}), we can know the non-interference
term of $\Gamma_{\cal M}(\l^{\prime-}\to l^-\bar{\nu}_i\nu_{i})$ equals to
$\Gamma_{\cal D}(\l^{\prime-}\to l^-\bar{\nu}_i\nu_{i})$.
Consequently, the partial decay width with the assumption of
Majorana neutrinos can be written as
\begin{eqnarray}
\Gamma_{\cal M}(l^{\prime-}\to l^-\bar{\nu}_l\nu_{l^{\prime}}) &=&\Gamma_{\cal D}(l^{\prime-}\to l^-\bar{\nu}_l\nu_{l^{\prime}})-\sum^{i=3}_{i=1}\Gamma^{int}_{\cal M}(\l^{\prime-}\to l^-\bar{\nu}_i\nu_{i})\,,
 \end{eqnarray}
where the $\Gamma^{int}_{\cal M}(\l^{\prime-}\to l^-\bar{\nu}_i\nu_{i})$ is the interference term
of $\Gamma_{\cal M}(\l^{\prime-}\to l^-\bar{\nu}_i\nu_{i})$ and is always positive. With
\begin{eqnarray}
  \Gamma_{\cal M}(l^{\prime-}\to l^-\bar{\nu}_l\nu_{l^{\prime}}) &<& \Gamma_{\cal D}(l^{\prime-}\to l^-\bar{\nu}_l\nu_{l^{\prime}})\,,
\end{eqnarray}
this interference term can be written as
\begin{eqnarray}\label{1a}
  \sum^{i=3}_{i=1}\Gamma^{int}_{\cal M}(\l^{\prime-}\to l^-\bar{\nu}_i\nu_{i}) &=&  \sum^{i=3}_{i=1}|U_{li}|^2|U^*_{l^{\prime}j}|^2 \frac{G^2_F}{2(2\pi)^3}\frac{1}{32m^3_{l^\prime}}\int\overline{| I^{int}_{\cal M}|}dm^2_{12} dm^2_{23}
\end{eqnarray}
and
\begin{eqnarray}\label{2a}
\overline{| I^{int}_{\cal M}|}&=&
\frac{1}{2}Tr[\gamma^\nu(\slashed{p}_3+m_3)\gamma^\mu(1-\gamma^5)v(p_1)\bar{v}(p_2)(1+\gamma^5)]\nonumber\\
&&\,\,Tr[\gamma_\mu(\slashed{p}_0+m_0)\gamma_\nu(1-\gamma^5)u(p_1)\bar{u}(p_2)(1+\gamma^5)]\,,
\end{eqnarray}
where
\begin{eqnarray}\label{3a}
  (1-\gamma^5)v(p_1)\bar{v}(p_2)(1+\gamma^5) &=& (1-\gamma^5)\slashed{p}_1\gamma^0\slashed{p}_2(1+\gamma^5)/(\sqrt{2E_1}\sqrt{2E_2})\nonumber\\
  (1-\gamma^5)u(p_1)\bar{u}(p_2)(1+\gamma^5) &=& (1-\gamma^5)\slashed{p}_1\gamma^0\slashed{p}_2(1+\gamma^5)/(\sqrt{2E_1}\sqrt{2E_2})\,.
\end{eqnarray}
Here, we have summed over the spin of $v(p_1)\bar{v}(p_2)$ and $u(p_1)\bar{u}(p_2)$,
and $E_1$ and $E_2$ are energies of neutrinos in the ${l^\prime}$ rest frame.
Using Eqs.~(\ref{1a}~,\ref{2a},~and~\ref{3a}), we can get
\begin{eqnarray}
   \sum^{i=3}_{i=1}\Gamma^{int}_{\cal M}(\l^{\prime-}\to l^-\bar{\nu}_i\nu_{i}) &=& \frac{G^2_F m^5_{l^{\prime}}}{24(2\pi)^3}Y_{ll^\prime}
\end{eqnarray}
and
\begin{eqnarray}
  Y_{ll^\prime} &=& \sum^{i=3}_{i=1}|U_{li}|^2|U^*_{l^{\prime}i}|^2\bigg\{(1-y_{ll^{\prime}})^2\ln y_{ll^{\prime}}\ln( 1-y_{ll^{\prime}})-6 y_{ll^{\prime}}(2-3y_{ll^{\prime}})\ln y_{ll^{\prime}}\nonumber\\
  &&-12(1-y_{ll^{\prime}})^2 Li_2(y_{ll^{\prime}})+2\pi^2(1-y_{ll^{\prime}})^2+(1-y_{ll^{\prime}})\frac{y^3_{ll^{\prime}}-11y^2_{ll^{\prime}}+61y_{ll^{\prime}}-39}{2}\bigg\}\,
\end{eqnarray}
where $Li_2$ is dilogarithm function~(Spence's function). Eventually, the ratio of the partial decay width
$\Gamma_{\cal M}(\tau^{-}\to \mu^-\bar{\nu}_\mu\nu_{\tau})$ to
$\Gamma_{\cal M}(\tau^{-}\to e^-\bar{\nu}_e\nu_{\tau})$ is determined to be
\begin{eqnarray}
  {\cal R_{M}} &=& \frac{\Gamma_{\cal M}(\tau^{-}\to \mu^-\bar{\nu}_\mu\nu_{\tau})}{\Gamma_{\cal M}(\tau^{-}\to e^-\bar{\nu}_e\nu_{\tau})}
  =\frac{K_{\mu\tau}-Y_{\mu\tau}}{K_{e\tau}-Y_{\mu\tau}}
\end{eqnarray}

\section{Numerical Results}
To perform the numerical analysis, the PMNS matrix elements as the theoretical inputs
 are given by 
 $(U_{e1},U_{e2},U_{e3})=(\cos\theta_{12}\cos\theta_{13},\sin\theta_{12}\cos\theta_{13},\sin\theta_{13}e^{i\delta})$,
 $(U_{\mu1},U_{\mu2},U_{\mu3})=(-\sin\theta_{12}\cos\theta_{23}-\cos\theta_{12}\sin\theta_{23}e^{i\delta},\cos\theta_{12}\cos\theta_{23}-\sin\theta_{12}\sin\theta_{23}\sin\theta_{13}e^{i\delta},\sin\theta_{23}\cos\theta_{13})$, and
$(U_{\tau1},U_{\tau2},U_{\tau3})=(\sin\theta_{12}\sin\theta_{23}-\cos\theta_{12}\cos\theta_{23}e^{i\delta},-\cos\theta_{12}\sin\theta_{23}-\sin\theta_{12}\cos\theta_{23}\sin\theta_{13}e^{i\delta},\nonumber\\
\cos\theta_{23}\cos\theta_{13})$
with $(\sin^2\theta_{12},\sin^2\theta_{23},\sin^2\theta_{13},\delta)=(0.307\pm0.013,0.546\pm0.021,0.0220\pm 0.0007,(1.36_{-0.16}^{+0.20})\pi)$~\cite{ParticleDataGroup:2020ssz} and
$(0<\theta_{12},\theta_{23},\theta_{13}<\pi/2)$~\cite{Xing:2020ijf}.
(We have ignored two CP-violating phases that are physically meaningful only if neutrinos are Majorana particles, which have no effect to all conclusions in this paper.)
Next, the data inputs are given by
\begin{eqnarray}
\sum^{i=3}_{i=1}|U_{ei}|^2|U^*_{\tau i}|^2&=&0.271\pm0.039,\;K_{e\tau}=1.0000\pm0.0000,\;Y_{e\tau}=0.065\pm0.009\nonumber\\
\sum^{i=3}_{i=1}|U_{\mu i}|^2|U^*_{\tau i}|^2&=&0.384\pm0.005,\;K_{\mu\tau}=0.9726\pm0.0000,\;Y_{\mu\tau}=0.090\pm0.001\nonumber\\
\sum^{i=3}_{i=1}|U_{e i}|^2|U^*_{\mu i}|^2&=&0.196\pm0.017,\;K_{e\mu}=0.9998\pm0.0000,\;Y_{e\mu}=0.047\pm0.004
\end{eqnarray}
and lifetime $t_\tau=(2.903\pm0.005)\times10^{-13}$~s and $t_\mu=(2.197\pm0.000)\times10^{-6}$~s.
Since Fermi constant $G_f$ comes from measurements of $\mu$ lepton lifetime from the decay channel
$\mu^{-}\to e^-\bar{\nu}_e\nu_{\mu}$, the data input of Fermi constant $G^{exp}_F=1.1664\times10^{-5}$~GeV$^{-2}$
from PDG can only be used for Dirac neutrino hypothesis $(G^{D}_F=G^{exp}_F)$.
Fermi constant for Majorana neutrino hypothesis are
$ G^{M}_F=G^{exp}_F\sqrt{\frac{K_e\mu}{K_{e\mu}-Y_{e\mu}}}=(1.1948\pm0.0025)\times10^{-5}$~GeV$^{-2}$.

All the calculated contrasts to experimental measurements are summarized in table~\ref{tyu}.
\begin{table}[b!]
\small
\caption{
Branching fraction (${\cal B}$)
and ${\cal R}$ for $\l^{\prime-}\to l^-\bar{\nu}_l\nu_{l^{\prime}}$.}\label{table1}
\setlength{\tabcolsep}{5mm}{
\begin{tabular}{|c|c|c|c|}
\hline
 $\l^{\prime-}\to l^-\bar{\nu}_l\nu_{l^{\prime}}$&Majorana neutrinos &Dirac neutrinos&experiment~\cite{ParticleDataGroup:2020ssz}\\
\hline\hline
${\cal B}(\tau^{-}\to \mu^-\bar{\nu}_\mu\nu_{\tau})\times 10^2$
&$16.53\pm 0.08$
&$17.36\pm 0.03$&$17.39\pm0.04$\\
${\cal B}(\tau^{-}\to e^-\bar{\nu}_e\nu_{\tau})\times 10^2$
&$17.51\pm 0.19$
&$17.85\pm0.03$&$17.82\pm0.04$\\
${\cal B}(\mu^{-}\to e^-\bar{\nu}_e\nu_{\mu})\times 10^2$
&$100.15\pm0.00$
&$100.15\pm0.00$&$\approx100$\\
${\cal R}=\frac{\Gamma(\tau^{-}\to \mu^-\bar{\nu}_\mu\nu_{\tau})}{\Gamma(\tau^{-}\to e^-\bar{\nu}_e\nu_{\tau})}$
&$0.9440\pm 0.0091$
&$0.9726\pm0.0000$&$0.9762\pm0.0028$\\
\hline
\end{tabular}\label{tyu}}
\end{table}

\section{Discussions and Conclusions}
As summarized in Table~\ref{tyu}, all the results under Dirac neutrino
hypothesis agrees with experimental results better than that of Majorana
neutrino hypothesis (except for the decay width
$\Gamma(\mu^{-}\to e^-\bar{\nu}_e\nu_{\mu})$, which is one of the inputs). In
particular, the ratio ${\cal R}$ of the decay width
$\Gamma(\tau^{-}\to \mu^-\bar{\nu}_\mu\nu_{\tau})$ to the decay width
$\Gamma(\tau^{-}\to e^-\bar{\nu}_e\nu_{\tau})$, which is related only to lepton
masses in Dirac neutrino hypothesis, while is involved with lepton masses and
neutrino mixing matrix parameters in Majorana neutrino hypothesis. The
uncertainties of theoretical calculations is comparable to that of experimental measurements.
The calculated result of ${\cal R}$ under Majorana neutrino hypothesis
deviates from the experimental measurements more than $3\sigma$, which is a strong
support to rule out Majorana neutrino hypothesis. However, the ${\cal R}$ value
under Dirac neutrino hypothesis is consistent with experimental measurements within
$1.3\sigma$, as well as the two branching fractions of $\tau^-$ decays.

All the discussion in this paper is based on the standard three-neutrino theory,
but even if we extend the standard three-neutrino model to other models, the
effect of identical particles will still contribute strong constraints.

\section{Summary And Outlook}
In summary, we have investigated $\l^{\prime-}\to l^-\bar{\nu}_l\nu_{l^{\prime}}$ decays.
According to the effect of identical particles,
we have found the difference between Dirac and Majorana neutrino hypotheses
and presented the comparisons to experimental measurements, indicating
a badly dislike to the hypothesis that neutrinos are Majorana particles and a strong
favor to the hypothesis that neutrinos are Dirac particles in the standard three-neutrino theory.

This work can be extended to study the angular distribution of polarization in purely
leptonic decays of leptons, which could give more crucial information to
distinguish Dirac neutrinos from Majorana neutrinos.
This is a focus of our next work and should also be the focus of future experimental
observations. For example, large and clean data sets of $e^+e^-\to\tau^+\tau^-$ have been
and will be collected at BESIII~\cite{Ablikim:2019hff}, and further study about polarization angles
of leptons in $\tau^-$ decay can be done.

\section*{ACKNOWLEDGMENTS}
YY was supported in part by  NSFC (Grant No.~11905023 and No.~12047564), the National Natural Science Foundation of China under (Grant  No.~12147102) and CQCSTC (Grants No.~cstc2020jcyj-msxmX0555).
BCK was supported in part by the Chinese Academy of Sciences (CAS) Large-scale Scientific Facility Program; Joint
Large-Scale Scientific Facility Fund of the NSFC and CAS (Contract No. U2032104) and NSFC (Grant No.~11875054 and No.~12192263).

\end{document}